\documentclass[twocolumn,letterpaper]{IEEEAerospaceCLS}  % only supports two-column, letterpaper format

\usepackage[]{graphicx}    % We use this package in this document
\usepackage[colorlinks=true, urlcolor=blue, linkcolor=black, citecolor=black]{hyperref} % ED 2025 added for web links
\usepackage[cmex10]{amsmath}
\usepackage{amssymb}
\usepackage{placeins}
\usepackage{algorithm}
\usepackage{algorithmic}
\newcommand{\ignore}[1]{}  % {} empty inside = %% comment
\hypersetup{%
pdftitle={A Data-Driven Surrogate Modeling and Sensor/Actuator Placement Framework for Flexible Spacecraft},
pdfauthor = {Matthew Hilsenrath and Daniel R. Herber}
}

\usepackage{layouts}

\begin{document}
\title{A Data-Driven Surrogate Modeling and Sensor/Actuator Placement Framework for Flexible Spacecraft}

\author{%
Matthew Hilsenrath\\ 
Colorado State University\\
Lockheed Martin\\
Evergreen, CO 80439\\
m.hilsenrath@colostate.edu
\and 
Daniel R. Herber\\
Colorado State University\\
% 200 W. Lake Street\\
Fort Collins, CO 80523\\
daniel.herber@colostate.edu
%%%% IMPORTANT: Use the correct copyright information--IEEE, Crown, or U.S. government. %%%%%
%\thanks{\footnotesize 979-8-3315-7360-7/26/$\$31.00$ \copyright2026 IEEE}              % This creates the copyright info that is the correct 2026 data.
}

\maketitle

\thispagestyle{plain}
\pagestyle{plain}

\begin{abstract}%
Flexible spacecraft structures present significant challenges for physical and control system design due to nonlinear dynamics, mission constraints, environmental variables, and changing operational conditions. This paper presents a data-driven framework for constructing reduced-order surrogate models of a flexible spacecraft using the method of Dynamic Mode Decomposition  (DMD), followed by optimal sensor/actuator pair placement. High-fidelity simulation data from a nonlinear flexible spacecraft model, including coupled rigid-body and elastic modes, are captured by defining a mesh of nodes over the spacecraft body. The data-driven methods are then used to construct a modal model from the time histories of these node points. Optimal sensor/actuator placement for controllability and observability is performed via a nonlinear programming technique that maximizes the singular values of the Hankel matrix. Finally, the sensor placement and dynamics modeling approach is iterated to account for changes in the dynamic system introduced by sensor/actuator physical mass. The proposed methodology enables initialization of physical modeling without requiring a direct analytical model and provides a practical solution for onboard implementation in model-based control and estimation systems. Results demonstrate optimal design methodology with substantial model-order reduction while preserving dynamic fidelity, and provide insight into effective sensor-actuator configurations for estimation and control.
\end{abstract} 
\tableofcontents
%%%%%%%%%%%%%%%%%%%%%%%%
\section{Introduction}
%%%%%%%%%%%%%%%%%%%%%%%%
Flexible spacecraft present significant modeling and control challenges due to coupled rigid-body and elastic dynamics, sensitivity to environmental inputs, and strict mission constraints. Classical analytical modeling techniques often fall short when applied to high-dimensional, nonlinear systems. High-fidelity finite element or modal models capture the dynamics, but can be computationally impractical. This gap has motivated strategies for tailored control design and modeling solutions alike.
Classical approaches rely on modal analysis to develop a dynamic system model, followed by controller augmentation. A direct example of this is input shaping coupled with proportional-derivative control for vibration attenuation in flexible solar arrays \cite{liu2016honeycomb}. Recently, model predictive controllers have been developed to handle actuator limits and disturbances that occur during thruster firings \cite{tracy2021mpc}. Data-driven methods are showing considerable potential across several areas of control design. Data-driven methods for flexible spacecraft predictive control are in development, which have the advantage of defining dynamics models directly from data, without the need to recognize and model the underlying physics \cite{wang2025deepc}.
Most of the existing work in this area is based on fixed sensor and actuator placement, neglecting the effects that the mass, location, and configuration alter the dynamic system under control in ways that may invalidate the controller or observer design, or at a minimum drive an optimized design out of the optimal configuration. Iterative loops that jointly update the reduced order model, optimize placement, and re-characterize dynamics remain largely unexplored.
This paper introduces a data-driven framework to iteratively construct reduced-order surrogate models, optimize sensor/actuator placement, and incorporate hardware into system dynamics. This approach yields a data-driven integrated approach to modeling, sensing, and actuation for the optimal design of controllable flexible spacecraft.
%%%%%%%%%%%%%%%%%%%%%%%%%%%%%%%%%%%%%%%%%%%%%%%%%%%%%%%
\section{Truth Model Simulation and Data}
\begin{figure}[tbh!]
\centering
\includegraphics[width=0.9\linewidth]{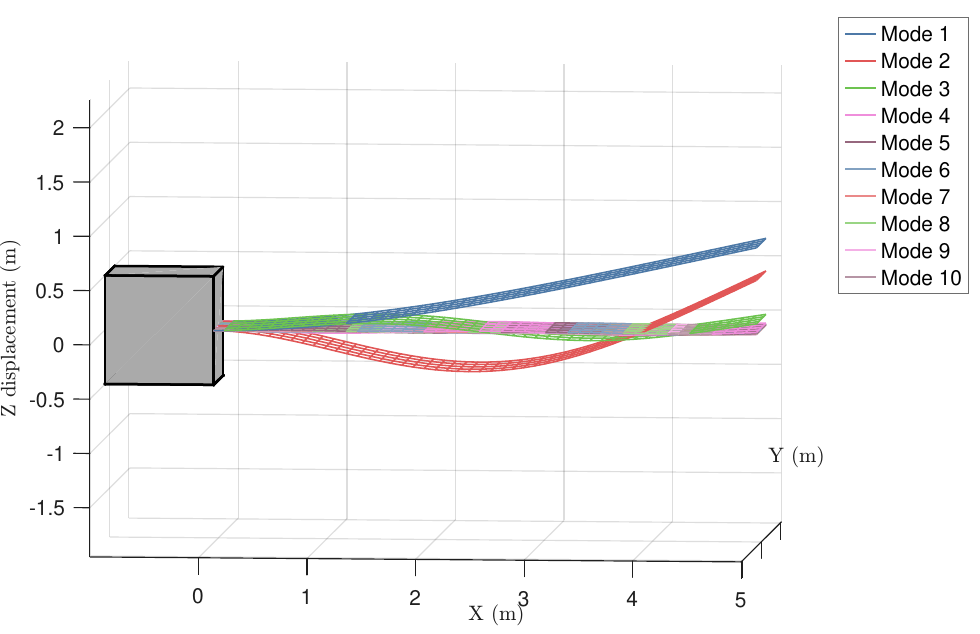}
\caption[10 modes]{\textbf{All 10 modes}}
\label{fig:all10modes}
\end{figure}

\begin{figure}[tbh!]
\centering
\includegraphics[width=0.8\linewidth]{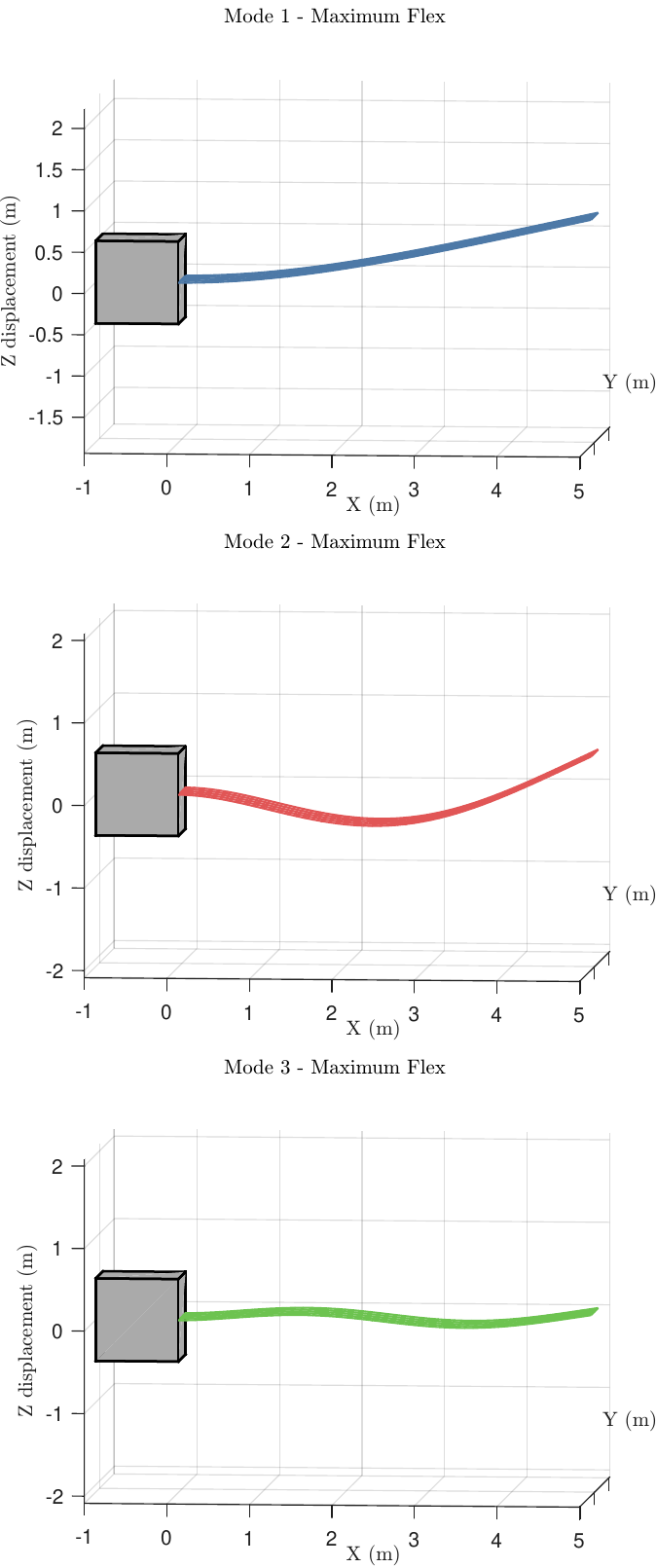}
\caption[10 modes]{\textbf{First 3 modes}}
\label{fig:first3modes}
\end{figure}

%%%%%%%%%%%%%%%%%%%%%%%%%%%%%%%%%%%%%%%%%%%%%%%%%%%%%%%
A nonlinear simulation of a flexible spacecraft structure is created in MATLAB to generate truth data. The spacecraft geometry includes rigid core dynamics and a flexible solar array appendage, discretized into a mesh of nodes. The simulation captures motion from 10 bending modes, derived from classical Euler-Bernoulli beam theory \cite{parks2004euler} with values from Table~\ref{tab:simvals} and illustrated in Figure~\ref{fig:all10modes}. To facilitate a clear demonstration of the method, amplitudes and damping characteristics were chosen to encourage dominant system behavior in the first three mode shapes, as illustrated in Figure~\ref{fig:first3modes}. Physical system dynamics typically result in lower-frequency modes that damp more slowly and higher-frequency modes that damp faster. 
\begin{table}[tbh]
\caption{\textbf{Truth model dynamic characteristics}} 
\centering 
\begin{tabular}{c c c c c} 
\hline\\[-1ex]
Mode  & Constant  & Amp & Freq.  & Damping \\
$i$ &  $\lambda$ & (m) &  $f_d$ (Hz) &  $\zeta$ \\[1ex]
\hline \\
1& 1.8751 & 0.800&    3.58  & 0.01\\[1ex]
2& 4.6941 & 0.500&  22.45  & 0.03\\[1ex]
3& 7.8548 & 0.100&  62.85  & 0.04\\[1ex]
4& 10.9955& 0.020&  122.85 & 0.08\\[1ex]
5& 14.1372& 0.010&  203.09& 0.08\\[1ex]
6& 17.2877& 0.010&  303.38& 0.08\\[1ex]
7& 20.4204& 0.005& 423.72& 0.08\\[1ex]
8& 23.5619& 0.005& 563.10& 0.10\\[1ex]
9& 26.7035& 0.002& 723.27& 0.10\\[1ex]
10&29.8451& 0.001& 903.47& 0.10\\[1ex]
% [1ex] adds vertical space
\hline % inserts single-line
\end{tabular}
\label{tab:simvals}
\end{table}

%%%%%%%%%%%%%%%%%%%%%%%%%%%%%%%%%%%%%%%%%%%%%%%%%%%%%%%%%%%%%%%%%
\section{Surrogate Modeling via DMD}
\label{sec:surrogateModel}
%%%%%%%%%%%%%%%%%%%%%%%%%%%%%%%%%%%%%%%%%%%%%%%%%%%%%%%%%%%%%%%%%
We apply Dynamic Mode Decomposition (DMD) to extract linear approximations of the spacecraft dynamics from high-dimensional time-series data. The goal is to construct a reduced-order model that captures the dominant modal behavior while accounting for actuator influence.

DMD is a data-driven technique for approximating the underlying linear dynamical system that best fits the evolution of observed states over time. Consider a discrete-time sequence of snapshots of a system state,
\begin{equation}
\mathbf{x}_1, \mathbf{x}_2, \dots, \mathbf{x}_m \in \mathbb{R}^n,
\end{equation}
taken at uniform time intervals. These snapshots are arranged into two data matrices,
\begin{align}
\mathbf{X} &= 
\begin{bmatrix}
	\mathbf{x}_1 & \mathbf{x}_2 & \cdots & \mathbf{x}_{m-1}
\end{bmatrix}
\in \mathbb{R}^{n \times (m-1)} \\
\mathbf{X}' &= 
\begin{bmatrix}
	\mathbf{x}_2 & \mathbf{x}_3 & \cdots & \mathbf{x}_m
\end{bmatrix}
\in \mathbb{R}^{n \times (m-1)}.
\end{align}
According to \cite{Tu2013} and \cite{Saito2020}, DMD fails in identifying standing wave modes, such as the vibrations present in an oscillating beam. For this case, the snapshot matrices may be appended with time-shifted snapshots,
\begin{align}
\mathbf{X} &= 
\begin{bmatrix}
	\mathbf{x}_1 & \mathbf{x}_2 & \cdots & \mathbf{x}_{m-2}\\
	\mathbf{x}_2 & \mathbf{x}_3 & \cdots & \mathbf{x}_{m-1}
\end{bmatrix} \\
\mathbf{X}' &= 
\begin{bmatrix}
	\mathbf{x}_2 & \mathbf{x}_3 & \cdots & \mathbf{x}_{m-1}\\
	\mathbf{x}_3 & \mathbf{x}_4 & \cdots & \mathbf{x}_{m}
\end{bmatrix},
\end{align}
which aids in standing wave identification.
We seek a best-fit linear operator $\mathbf{A} \in \mathbb{R}^{n \times n}$ such that
\begin{equation}
\mathbf{X}' \approx \mathbf{A} \mathbf{X}.
\end{equation}
The DMD projects this operator onto a low-rank subspace using Singular Value Decomposition (SVD). Compute the reduced SVD of $\mathbf{X}$ as
\begin{equation}
\mathbf{X} = \mathbf{U}_r \boldsymbol{\Sigma}_r \mathbf{V}_r^\top,
\end{equation}
where $\mathbf{U}_r \in \mathbb{R}^{n \times r}$, $\boldsymbol{\Sigma}_r \in \mathbb{R}^{r \times r}$, and $\mathbf{V}_r \in \mathbb{R}^{(m-1) \times r}$ retain only the leading $r$ singular components. The low-dimensional approximation of $\mathbf{A}$ is then given by
\begin{equation}
\tilde{\mathbf{A}} = \mathbf{U}_r^\top \mathbf{X}' \mathbf{V}_r \boldsymbol{\Sigma}_r^{-1} \in \mathbb{R}^{r \times r}.
\end{equation}
The eigendecomposition of $\tilde{\mathbf{A}}$,
\begin{equation}
\tilde{\mathbf{A}} \mathbf{W} = \mathbf{W} \boldsymbol{\Lambda},
\end{equation}
yields DMD eigenvalues $\boldsymbol{\Lambda} = \operatorname{diag}(\lambda_1, \dots, \lambda_r)$ and eigenvectors $\mathbf{W}$. The corresponding dynamic modes in the original state space are
\begin{equation}
\boldsymbol{\Phi} = \mathbf{X}' \mathbf{V}_r \boldsymbol{\Sigma}_r^{-1} \mathbf{W} \in \mathbb{R}^{n \times r}.
\end{equation}
Each column $\boldsymbol{\phi}_j$ of $\boldsymbol{\Phi}$ is a DMD mode, and each eigenvalue $\lambda_j$ governs its temporal evolution via $\lambda_j^t$. Defining initial modal amplitudes $\mathbf{b} \in \mathbb{C}^r$, the time-evolution of the system can be reconstructed as
\begin{equation}
\mathbf{x}_t \approx \sum_{j=1}^r \boldsymbol{\phi}_j \lambda_j^{t-1}b_j , \qquad t = 1, 2, \dots
\end{equation}
The eigenvalues $\lambda_j$ encode both growth/decay (via $|\lambda_j|$) and oscillatory frequency (via $\mathrm{Im}(\lambda_j)$). This decomposition yields a representative description of the system’s dynamics in terms of spatial modes and their temporal evolution, enabling reconstruction and analysis in both time and frequency domains.

\begin{figure}[tbh!]
\centering
\includegraphics[scale=0.55]{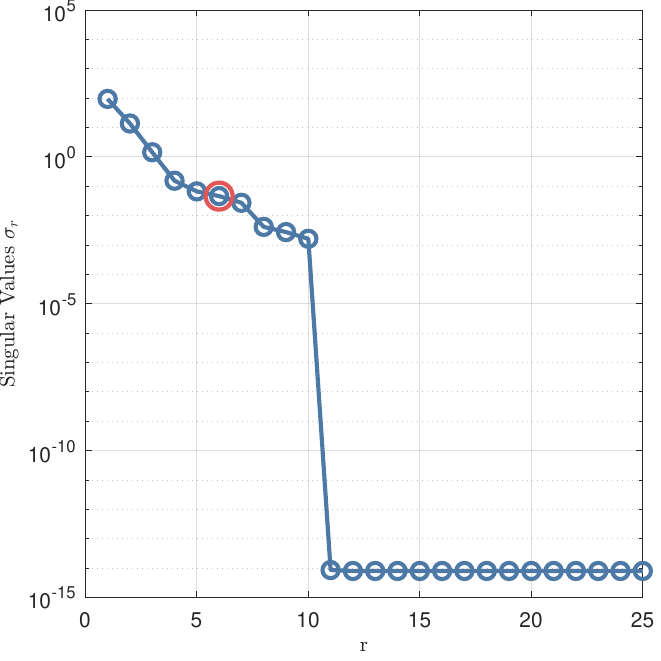}
\caption{\textbf{Singular value decomposition of truth model output data}}
\label{fig:svd1}
% \end{figure}
% \begin{figure}[tbh]
\centering
\includegraphics[scale=0.55]{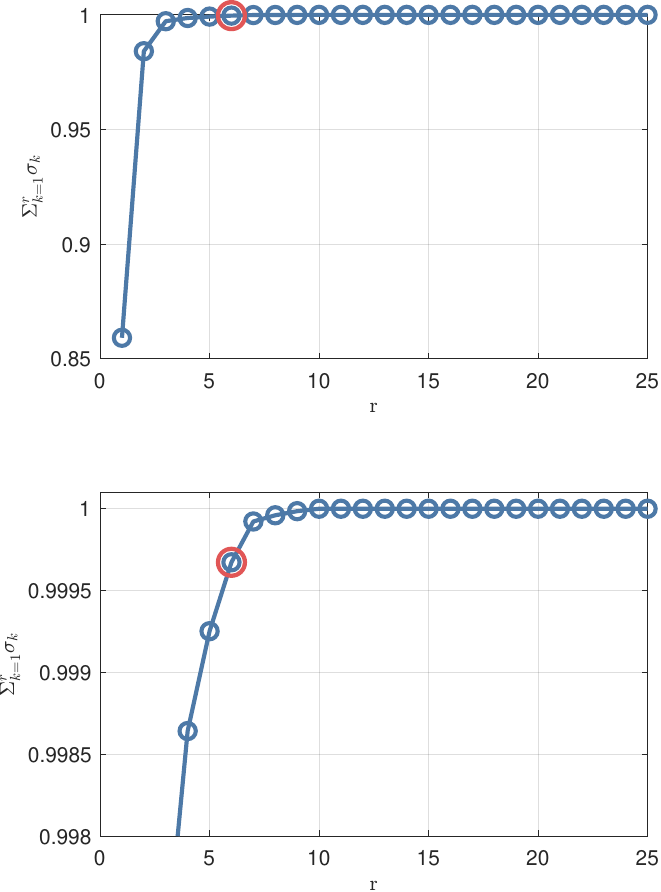}
\caption{\textbf{Cumulative sum, singular value decomposition of truth model}}
\label{fig:svd2}
\end{figure}

\begin{figure}[tbh!]
\centering
\includegraphics[width=0.85\linewidth]{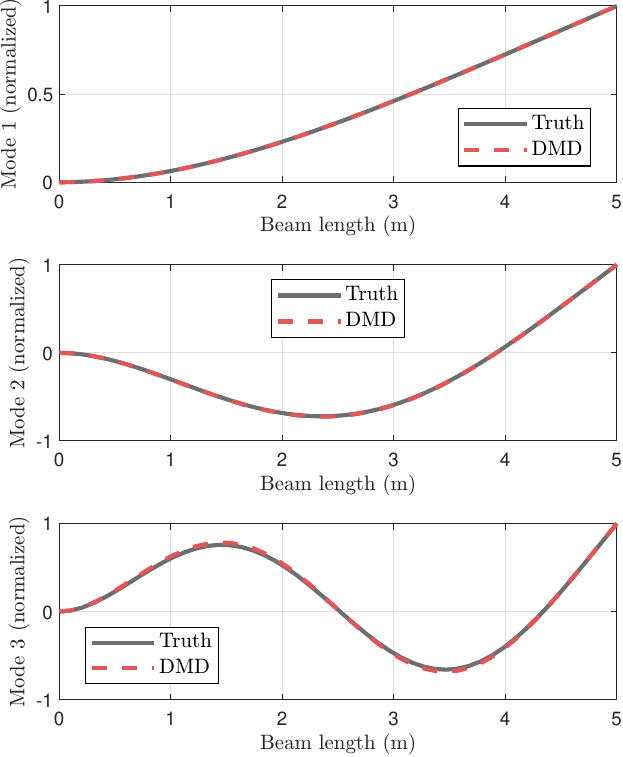}
\caption{\textbf{Dominant 3 mode shapes, normalized DMD reconstruction overlaying normalized truth mode}}
\label{fig:mds}
\end{figure}

\begin{figure}[tbh!]
\centering
\includegraphics[scale=0.55]{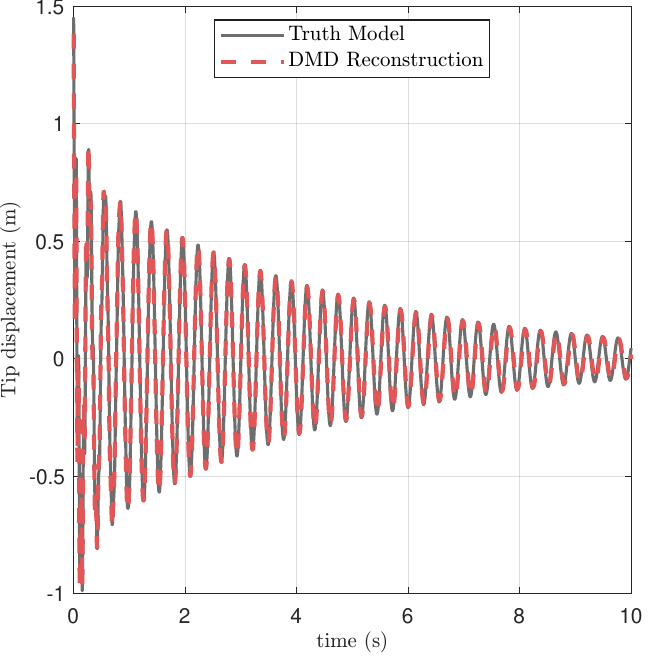}
\caption{\textbf{Time domain DMD reconstruction of tip response overlaying truth tip response} }
\label{fig:tip}
% \end{figure}
% \begin{figure}[tbh]
\centering
\includegraphics[scale=0.55]{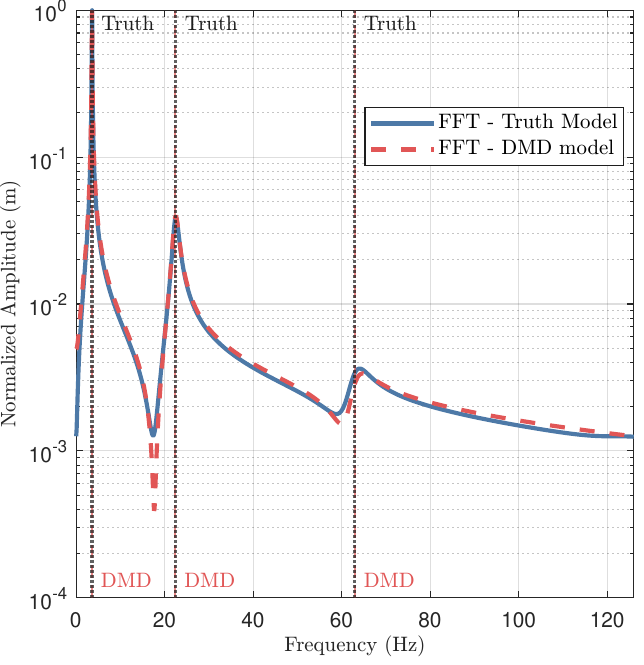}
\caption{\textbf{Frequency domain representation of truth model and DMD reconstruction.}}
\label{fig:fft}
\end{figure}

The application of SVD on the time-domain truth model data is shown in Figure~\ref{fig:svd1} and Figure~\ref{fig:svd2}. Figure~\ref{fig:svd1} shows the first 10 singular values significantly higher than subsequent values, indicating greater contribution to the overall dynamics of the system. Figure~\ref{fig:svd2} shows the normalized cumulative sum of the singular values, indicating the total amount of energy of the dynamic system. In both subfigures of Figure~\ref{fig:svd2}, the red marker signifies the 6th singular value. Since we intend to determine a surrogate model of the system truncated to the first 3 dynamic modes, the first 6 singular values are extracted (to account for complex conjugate pairs). Figure~\ref{fig:svd2} shows that more than 99.95\% of system energy is captured in the first three modes.
The reduced-order model constructed from the results of DMD captures the time domain as well as frequency domain characteristics of the original system. Figures \ref{fig:mds}, \ref{fig:tip}, and \ref{fig:fft} show agreement between the dynamics driving the truth model and the data-driven reconstruction produced by the DMD method. 
%%%%%%%%%%%%%%%%%%%%%%%%%%%%%%%%%%%%%%%%%%%%%%%%%%%%%%%%%%%%%%%%%%%%%%%%%%%%%%
\section{Equivalence of the Data-Driven Hankel and the Gramian Product Hankel}
\label{sec:hankel_equivalence}
%%%%%%%%%%%%%%%%%%%%%%%%%%%%%%%%%%%%%%%%%%%%%%%%%%%%%%%%%%%%%%%%%%%%%%%%%%%%%%
\subsection{Problem Statement and Assumptions}
In order to optimally place sensor/actuator pairs, it is necessary to define a balanced representation of the controllability and observability Gramians. This representation is clearly defined for a linear time-invariant (LTI) system. When producing a data-defined system, properties of the Hankel matrix and its singular value decomposition are recognized, which allow a proof that the balanced observability and controllability Gramians can be produced from the data-driven Hankel matrix.
To produce this proof, consider a discrete-time LTI system in state-space form,
\begin{align}
	\mathbf{x}_{k+1} &= \mathbf{A} \mathbf{x}_k + \mathbf{B} \mathbf{u}_k, \qquad \mathbf{x}_0 = 0, \label{eq:state}\\
	\mathbf{y}_k &= \mathbf{C} \mathbf{x}_k,\label{eq:output}
\end{align}
where \(\mathbf{A}\in\mathbb{R}^{n\times n},\; \mathbf{B}\in\mathbb{R}^{n\times m},\; \mathbf{C}\in\mathbb{R}^{p\times n}\).
We assume the system is \emph{asymptotically stable} (all eigenvalues of \(\mathbf{A}\) inside the unit circle), so the infinite-horizon Gramians exist.
Next, the \emph{Markov parameters} (impulse-response matrices) are defined as
\begin{equation}
h_k \;=\; \mathbf{C} \mathbf{A}^{k-1} \mathbf{B} \in \mathbb{R}^{p\times m}, \qquad k=1,2,\dots
\end{equation}
which are the discrete-time output samples in response to an impulse input at time \(0\) (initial iterations shown in Table~\ref{tab:markov}). The corresponding Hankel matrix will also require a defined number of rows and columns. 
\begin{table}[ht]
\centering % centering table
\caption{Example Markov parameter iterations}
\label{tab:markov}
\begin{tabular}{l c c} % creating 10 columns
$k$ & LTI iteration & Markov parameter
\\ [0.5ex]
\hline \\ [0.5ex]% inserts single-line
% Entering 1st row
& $\mathbf{x}_1 = \mathbf{A}(0)+\mathbf{B}(1) = \mathbf{B}$ &$=h_1$ \\[-1ex]
\raisebox{1.5ex}{$1$}  &$\mathbf{y}_0 = \mathbf{C}(\mathbf{B})$
&  \\[1ex]
% Entering 2nd row
& $\mathbf{x}_2 = \mathbf{A}(\mathbf{B})+\mathbf{B}(0) = \mathbf{AB}$ &$=h_2$ \\[-1ex]
\raisebox{1.5ex}{$2$}  &$\mathbf{y}_1 = \mathbf{C}(\mathbf{AB})$
&  \\[1ex]
% Entering 3rd row
& $\mathbf{x}_3 = \mathbf{A}(\mathbf{AB})+\mathbf{B}(0) = \mathbf{A}^2\mathbf{B}$ &$=h_3$ \\[-1ex]
\raisebox{1.5ex}{$3$}  &$\mathbf{y}_2 = \mathbf{C}(\mathbf{A}^2\mathbf{B})$
&  \\[1ex]
% [1ex] adds vertical space
\hline % inserts single-line
\end{tabular}
\label{tab:PPer}
\end{table}
The parameters $r$ and $s$, which define the number of columns and rows in the Hankel matrix, are chosen from the data by estimating the dominant temporal period of the measured signal. The slowest oscillatory mode present in the data ultimately defines these parameters to ensure all characteristic dynamics are present, without unnecessary burden from excessive computation or ill-conditioned matrices. The final choice of $s$ and $r$ balances capturing the dominant oscillation dynamics while maintaining a well-conditioned Hankel structure.

After we choose integers \(s\ge 1\) (number of block rows) and \(r\ge 1\) (number of columns), we may form the block Hankel matrix \(\mathcal{H}_0\in\mathbb{R}^{(s p)\times (r m)}\) from Markov parameters:
\begin{equation}\label{eq:hankel_markov}
	\mathcal{\mathbf{H}}_0 \;=\;
	\begin{bmatrix}
		h_1 & h_2 & \cdots & h_r \\
		h_2 & h_3 & \cdots & h_{r+1} \\
		\vdots & \vdots & \ddots & \vdots \\
		h_s & h_{s+1} & \cdots & h_{s+r-1}
	\end{bmatrix}.
\end{equation}
\subsection{Similarity to Observability and Controllability Matrices}
Define the finite (block) observability matrix \( \mathcal{O}_s \in \mathbb{R}^{(s p)\times n} \) and the finite (block) controllability matrix \( \mathcal{C}_r \in \mathbb{R}^{n\times (r m)} \) by
\begin{align}
	\mathcal{O}_s &\;=\;
	\begin{bmatrix}
		\mathbf{C} \\[4pt]
		\mathbf{CA} \\[4pt]
		\vdots \\[4pt]
		\mathbf{CA}^{\,s-1}
	\end{bmatrix} \\
	\mathcal{C}_r &\;=\; 
	\begin{bmatrix}
		\mathbf{B} & \mathbf{AB} & \mathbf{A}^2\mathbf{B} & \cdots & \mathbf{A}^{\,r-1}\mathbf{B}
	\end{bmatrix}.
\end{align}
A direct multiplication yields the Hankel factorization (the Ho-Kalman/ERA identity from \cite{juang1985pappa}):
\begin{align}\label{eq:hankel_factor}
	\mathcal{O}_s \, \mathcal{C}_r &= \begin{bmatrix}
		\mathbf{C} \\[4pt]
		\mathbf{CA} \\[4pt]
		\vdots \\[4pt]
		\mathbf{CA}^{\,s-1}
	\end{bmatrix}\begin{bmatrix}
	\mathbf{B} & \mathbf{AB} & \cdots & \mathbf{A}^{\,r-1}\mathbf{B}
	\end{bmatrix}\\
	&\;=\; \begin{bmatrix}
		\mathbf{CB} & \mathbf{CAB} &  \cdots & \mathbf{CA}^{r-1}\mathbf{B}\\
		\mathbf{CAB} & \mathbf{CA}^2\mathbf{B} &  \cdots& \mathbf{CA}^{r}\mathbf{B}\\
		\vdots& \vdots & \ddots& \vdots\\
		\mathbf{CA}^{s-1}\mathbf{B} & \mathbf{CA}^s\mathbf{B} &  \hdots& \mathbf{CA}^{s+r-2}\mathbf{B}\\
	\end{bmatrix} \\
	&\;=\;\begin{bmatrix}
		h_1 & h_2 & \cdots & h_r \\
		h_2 & h_3 & \cdots & h_{r+1} \\
		\vdots & \vdots & \ddots & \vdots \\
		h_s & h_{s+1} & \cdots & h_{s+r-1}
	\end{bmatrix}\\
	&\;=\; \mathcal{\mathbf{H}}_0 
\end{align}
which shows the \((s,r)\)-th block of \(\mathcal{O}_s\mathcal{C}_r\) is \(\mathbf{C} \mathbf{A}^{s-1} \mathbf{A}^{r-1} \mathbf{B} = \mathbf{C} \mathbf{A}^{s+r-2} \mathbf{B}\). This matches the \((s,r)\)-th block \(h_{s+r-1}\) of the Hankel matrix in Eq.~\eqref{eq:hankel_markov}, showing that the data-driven Hankel matrix is equivalent to the outer-product of observability and controllability matrices. 
\subsection{Finite Approximations to the Gramians}
Define the finite (truncated) controllability and observability Gramian approximations
\begin{align}
\mathbf{W}_{c}^{(r)} &\;=\; \sum_{k=0}^{r-1} \mathbf{A}^k \mathbf{B} \mathbf{B}^\top (\mathbf{A}^k)^\top \;=\; \mathcal{C}_r \mathcal{C}_r^\top \;\in\mathbb{R}^{n\times n}, \label{eq:Wc_trunc}\\[4pt]
\mathbf{W}_{o}^{(s)} &\;=\; \sum_{k=0}^{s-1} (\mathbf{A}^k)^\top \mathbf{C}^\top \mathbf{C} \mathbf{A}^k \;=\; \mathcal{O}_s^\top \mathcal{O}_s \;\in\mathbb{R}^{n\times n}. \label{eq:Wo_trunc}%
\end{align}
If the system is stable, as \(r\to\infty\) and \(s\to\infty\), these truncated sums converge to the infinite-horizon Gramians
\begin{align}
\mathbf{W}_c \;&=\; \sum_{k=0}^{\infty} \mathbf{A}^k \mathbf{B} \mathbf{B}^\top (\mathbf{A}^k)^\top\\
\mathbf{W}_o \;&=\; \sum_{k=0}^{\infty} (\mathbf{A}^k)^\top \mathbf{C}^\top \mathbf{C} \mathbf{A}^k
\end{align}
Now if we compute the singular value decomposition of \(\mathcal{\mathbf{H}}_0\):
\begin{equation}\label{eq:svd_H}
	\mathcal{\mathbf{H}}_0 \;=\; \mathbf{U} \mathbf{\Sigma} \mathbf{V}^\top,
\end{equation}
with \(\mathbf{\Sigma}=\mathrm{diag}(\sigma_1,\dots,\sigma_q)\), \(\sigma_1\ge\sigma_2\ge\cdots\ge 0\), and \(q=\operatorname{rank}(\mathcal{\mathbf{H}}_0)\), we obtain
\begin{align}
	\mathcal{\mathbf{H}}_0 \mathcal{\mathbf{H}}_0^\top
	\;=\; \mathcal{O}_s \mathcal{C}_r \mathcal{C}_r^\top \mathcal{O}_s^\top
	\;=\; \mathcal{O}_s \, W_c^{(r)} \, \mathcal{O}_s^\top,
	\label{eq:HHT}
\end{align}
and similarly
\begin{align}
	\mathcal{\mathbf{H}}_0^\top \mathcal{\mathbf{H}}_0
	\;=\; \mathcal{C}_r^\top \mathcal{O}_s^\top \mathcal{O}_s \mathcal{C}_r
	\;=\; \mathcal{C}_r^\top \, W_o^{(s)} \, \mathcal{C}_r.
	\label{eq:HTH}
\end{align}
By properties of the SVD, the nonzero singular values \(\{\sigma_i\}\) satisfy
\begin{align}
	\sigma\{\mathcal{\mathbf{H}}_0\mathcal{\mathbf{H}}_0^\top\} &= \sigma\{\mathcal{\mathbf{H}}_0^\top\mathcal{\mathbf{H}}_0\}\\
	&= \sigma\{\mathcal{O}W_c\mathcal{O}^\top\}\\
	&= \sigma\{\mathcal{C}^\top W_o\mathcal{C}\}
\end{align}
Therefore, the Hankel singular values (HSVs), $\sqrt{\sigma\{\mathcal{\mathbf{H}}_0\mathcal{\mathbf{H}}_0^\top\}}$, converge to the balanced realization of the controllability and observability Gramians, $\hat{\mathbf{W}}_o$ and $\hat{\mathbf{W}}_c$, which are the eigenvalues of these Gramians, but produced strictly from data ($\mathbf{y}_k$). This establishes the equivalence of the two Hankel constructions in the context of stable LTI systems and sufficiently rich (impulse-like) data. The \emph{data-driven Hankel}  and the \emph{model-based Hankel} are the same object in the sense that their singular-value spectra coincide under the standard ERA limits. Note that the proof demonstrates equality of \emph{spectra} (nonzero eigenvalues), not matrix equality. In general
\[
\mathcal{\mathbf{H}}_0\mathcal{\mathbf{H}}_0^\top \;=\; \mathcal{O}_s W_c^{(r)} \mathcal{O}_s^\top
\ \text{and}\ 
\mathbf{W}_c^{(r)} \mathbf{W}_o^{(s)} \;=\; \mathcal{C}_r \mathcal{C}_r^\top \mathcal{O}_s^\top \mathcal{O}_s.
\]
This justifies using the data-driven Hankel singular values directly in the following placement objective.
%%%%%%%%%%%%%%%%%%%%%%%%%%%%%%%%%%%%%%%%%%%%%%%%
\section{Sensor/Actuator Placement Optimization}
%%%%%%%%%%%%%%%%%%%%%%%%%%%%%%%%%%%%%%%%%%%%%%%%
Following data-driven system identification by Dynamic Mode Decomposition (DMD), optimal sensor/actuator pair locations are determined through an iterative optimization framework. The DMD approach yields a data-driven, reduced-order LTI approximation of the underlying solar array flexible mode dynamics. The optimization criterion employs the cost function proposed by \cite{Maghami1993}, which maximizes the intersection of system controllability and observability subspaces for optimal sensor/actuator placement.

A fundamental challenge arises from the bidirectional coupling between sensor/actuator placement and system dynamics. The addition of sensor/actuator mass perturbs the structural dynamic properties, particularly the modal characteristics of the flexible array. Consequently, placement locations influence the system dynamics, while the modified dynamics, in turn, affect the optimal placement solution. The main contribution that this paper presents is a novel approach that leverages the data-driven Hankel matrix representation obtained from DMD analysis to iteratively converge upon optimal sensor/actuator placement by minimizing the Maghami-Joshi cost function:%
\begin{align}%
\min_{\mathbf{x}}:\; \quad & J(\mathbf{x}) \;=\; \sum_{i=1}^{n_s}\frac{1}{\mathbf{\sigma}_i\big(\mathbf{H(x)}\big)}\\[4pt]
\text{subject\;to}:\; \quad & \mathbf{x}^{L} \le \mathbf{x} \le \mathbf{x}^{U}
\end{align}
\subsection{Optimal Sensor Placement via Hankel Singular Values}
The eigenvectors corresponding to the dominant modes extracted via DMD can be used to reconstruct the dominant system dynamics of any nodes along the array \cite{kutz2016dynamic}, as shown in Section~\ref{sec:surrogateModel}: 
\begin{equation}
	\mathbf{Y}_{\text{DMD}}(t) = \sum_{j=1}^{n_r} \text{Re}\left\{\boldsymbol{\phi}_j \lambda_j^t b_j\right\}, \quad t = 1, 2, \ldots, n_t-1
	\label{eq:dmd_reconstruction}
\end{equation}
where:
\begin{itemize}
	\item $\mathbf{Y}_{\text{DMD}} \in \mathbb{R}^{n_x \times n_t}$ is the reconstructed data matrix
	\item $\boldsymbol{\phi}_j \in \mathbb{C}^{n_x}$ represents the $j$-th DMD mode (eigenvector of the DMD operator)
	\item $\lambda_j \in \mathbb{C}$ is the $j$-th eigenvalue corresponding to the $j$-th DMD mode
	\item $b_j \in \mathbb{C}$ is the $j$-th initial amplitude coefficient
	\item $n_r$ is the number of selected dominant modes used for reconstruction
	\item $n_x$ is the number of spatial locations (nodes along the array)
	\item $n_t$ is the total number of time snapshots
\end{itemize}
Our goal is to select $n_a$ sensors from a set of candidates $\mathcal{C} \subseteq \{1, \dots, p\}$ such that the chosen sensors capture the most dynamic information about the system.
We construct a \textit{block Hankel matrix} from the measurements of each candidate sensor subset.  
The singular values of this matrix measure how much independent dynamic content is captured: large singular values correspond to well-observed system dynamics.  
Thus, we are justified in using a cost function that penalizes small singular values:
\begin{equation}
J(\mathcal{C}) = \sum_{i=1}^{n_r} \frac{1}{\sigma_i(\mathbf{H(\mathcal{C})})},
\end{equation}
where $\sigma_i(\mathbf{H(\mathcal{C})})$ are the singular values of the Hankel matrix for candidate sensor set $\mathcal{C}$ and $n_r$ is the truncation index, chosen for the selected number of dominant modes.  Minimizing $J(\mathcal{C})$ selects sensor locations that best capture the dominant dynamics, avoiding sets that miss important modes.

\begin{figure}[p!]
	\centering
	\includegraphics[scale=0.6]{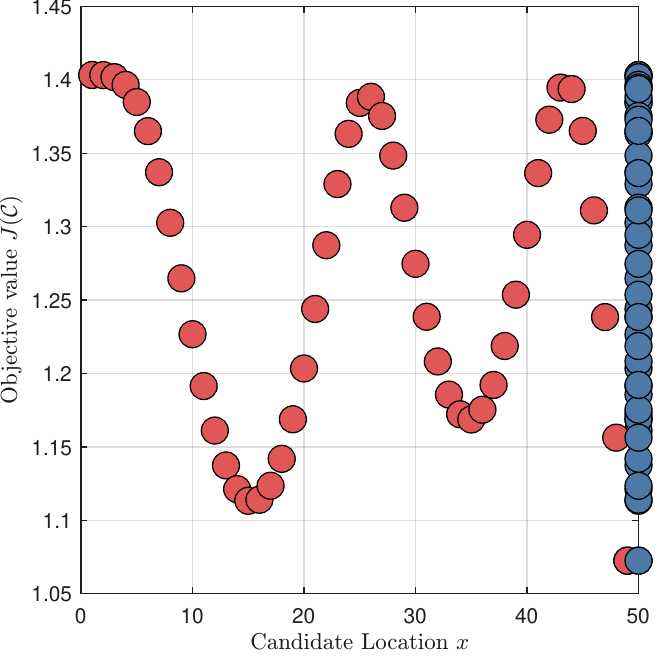}
	\caption{\textbf{Objective function values over candidate locations of unloaded beam}}
	\label{fig:cand1}
% \end{figure}

% \begin{figure}[tbhp!]
	\centering
	\includegraphics[width=0.9\linewidth]{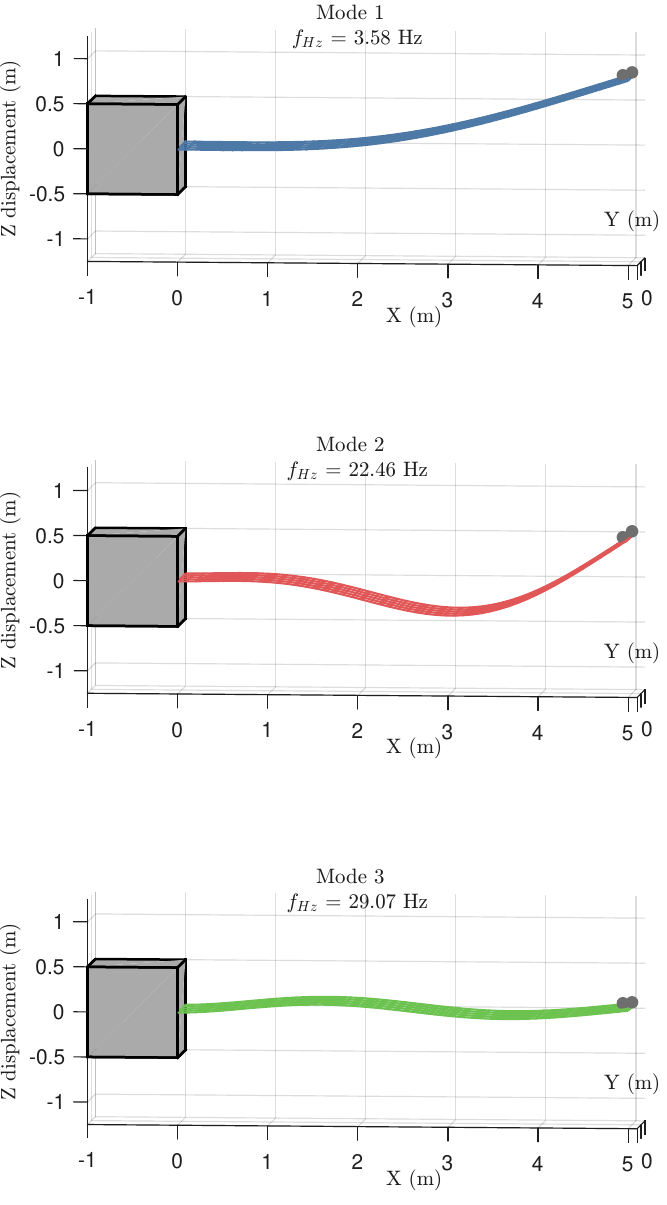}
	\caption{\textbf{Na\"ive optimal placement (suboptimal placement) of sensor/actuator pairs (this placement does not account for the mass of the sensor/actuator pairs)}}
	\label{fig:ANC1}
\end{figure}

For a subset of sensors $\mathcal{C}$, the block Hankel matrix for $\mathbf{H} \in \mathbb{R}^{(sk) \times r}$ is defined for each data point $\mathbf{y}_i(t) \in \mathbf{Y}_{\text{DMD}}$ as
\begin{equation}
\mathbf{H} = \begin{bmatrix}
		\mathbf{y}_1(1) & \mathbf{y}_1(2) & \cdots & \mathbf{y}_1(r) \\
		\mathbf{y}_1(2) & \mathbf{y}_1(3) & \cdots & \mathbf{y}_1(r+1) \\
		\vdots & \vdots & \ddots & \vdots \\
		\mathbf{y}_1(s) & \mathbf{y}_1(s+1) & \cdots & \mathbf{y}_1(s+r-1) \\[0.3em]
		\mathbf{y}_2(1) & \mathbf{y}_2(2) & \cdots & \mathbf{y}_2(r) \\
		\vdots & \vdots & \ddots & \vdots \\
		\mathbf{y}_2(s) & \mathbf{y}_2(s+1) & \cdots & \mathbf{y}_2(s+r-1) \\[0.3em]
		\vdots & \vdots & \ddots & \vdots \\[0.3em]
		\mathbf{y}_{k}(1) & \mathbf{y}_{k}(2) & \cdots & \mathbf{y}_{k}(r) \\
		\vdots & \vdots & \ddots & \vdots \\
		\mathbf{y}_{k}(s) & \mathbf{y}_{k}(s+1) & \cdots & \mathbf{y}_{k}(s+r-1)
\end{bmatrix}
\end{equation}
where:
\begin{itemize}
	\item $y_i(t)$ represents the $i^{th}$ selected output signal from $\mathcal{C}$ at time $t$ 
	\item $s$ is the number of block rows per output
	\item $k$ is the total number of selected outputs
	\item $r$ is the $\mathrm{col}(\mathcal{C}) - 2s + 1$ (number of columns)
\end{itemize}
 To exercise the effectiveness of the method, a selection space was chosen where a combinatorial search is computationally feasible. Evaluation of the cost function over all possible sensor/actuator pair locations for 2 sensors yields a quadratic complexity search space $O(n^2)$. The sensor/actuator selection optimization is outlined in Algorithm~\ref{alg:bruteforce}. 
\begin{algorithm}[b]
\caption{Initial sensor location selection based on Hankel singular values}
\label{alg:bruteforce}
\begin{algorithmic}[1]
\REQUIRE Measurement matrix $\mathbf{Y}_{\text{DMD}} \in \mathbb{R}^{n_x \times n_t}$, candidate sensors $\mathcal{C}$, number of sensors $n_a$, Hankel depth $s$, truncation index $n_r$
\ENSURE Optimal sensor locations $\mathcal{C}^\star$ by minimizing cost $J(\mathcal{C}^\star)$
\STATE Generate all possible subsets of $n_a$ sensors from $\mathcal{C}$
\FOR{each candidate subset $\mathcal{C}$}
\STATE Build block Hankel matrix $\mathbf{H}(\mathcal{C})$ using the selected sensors
\STATE Compute singular values $\sigma_1 \ge \dots \ge \sigma_{n_r}$ of $\mathbf{H}(\mathcal{C})$
\STATE Evaluate cost function $J(\mathcal{C}) = \sum_{i=1}^{n_r} 1 / \sigma_i$ % Maghami-Joshi
\ENDFOR
\STATE Select subset $\mathcal{C}^\star$ with minimum cost $J(\mathcal{C}^\star)$
\STATE $\mathcal{C}^\star$, $J(\mathcal{C}^\star)$
\end{algorithmic}
\end{algorithm}
Algorithm objective function values are plotted in Figure~\ref{fig:cand1}. The evaluation was performed using an inner and outer loop to ensure an exhaustive search, and the sensor locations corresponding to the inner loop's minimum objective quantities are shown in the plot. Each objective function value is marked with two corresponding candidate locations, in red and blue. The inner loop begins with a candidate location in red and determines the corresponding candidate location that will minimize the objective function in blue. The next iteration is the adjacent candidate location to the previous position, and its corresponding candidate position, which minimizes the objective function. For this structure, the corresponding location that minimizes any other candidate location happens to be the tip of the beam; hence, the string of blue points carries down the objective values at candidate location 50. The resulting sensor/actuator placement is illustrated over the dominant 3 mode shapes in Figure~\ref{fig:ANC1}.
\subsection{Modified Dynamics with Sensor/Actuator Placement}
Placement of sensor/actuator pairs alters the beam dynamics due to their nonzero mass. The dynamic behavior of a cantilever beam with attached point masses is modeled using the analytical-and-numerical-combined (ANC) method~\cite{wu1990anc}. This approach combines closed-form expressions for unloaded beam modes with numerical eigenvalue correction due to lumped masses.

When point masses $m_j$ are introduced at positions $x_j$, the mass-perturbed eigenproblem takes the form
\begin{equation}\label{eq:ANC}
	\left( \mathbf{I} + \sum_{j=1}^{p} m_j \, \phi(x_j)\phi^\top(x_j) \right)\boldsymbol{\eta} 
	= \frac{\omega^2}{\bar{\omega}^2} \, \boldsymbol{\eta},
\end{equation}
where $\bar{\omega}$ denotes the corrected natural frequencies and $\boldsymbol{x}$ the modal participation factors. The corrected mode shapes are then reconstructed as linear combinations of the unloaded modes, scaled by $\boldsymbol{\eta}$.
This formulation was implemented to evaluate the first three natural frequencies and associated mode shapes for the solar array with two point masses representing the sensor/actuator pairs.  
%%%%%%%%%%%%%%%%%%%%%%%%%%%%%%%%%%%%%%%%%%%%%%
\subsection{Algorithm Implementation}
\begin{figure}[p!]
	\centering
	\includegraphics[scale=0.6]{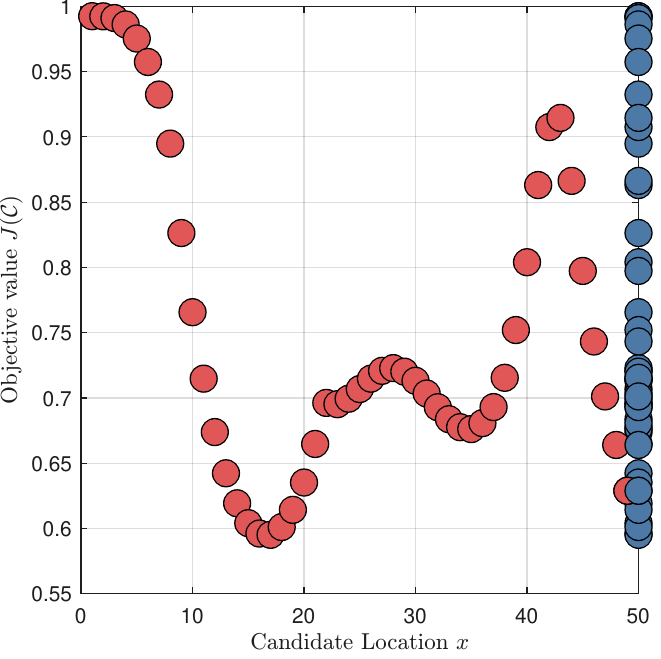}
	\caption{\textbf{Objective function values over candidate location of beam with mass properties of sensor/actuator pairs}}
	\label{fig:cand2}
% \end{figure}
% \begin{figure}[tbhp!]
	\centering
	\includegraphics[width=0.9\linewidth]{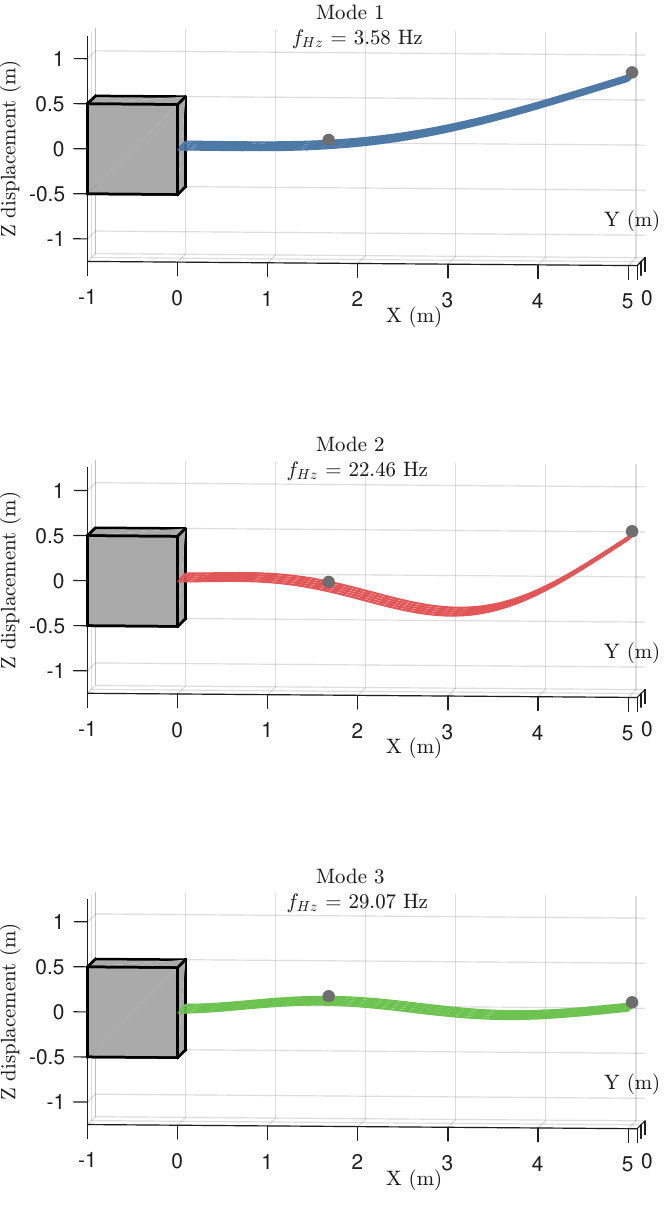}
	\caption{\textbf{Mass-adjusted optimal placement}}
	\label{fig:ANC2}
\end{figure}

%%%%%%%%%%%%%%%%%%%%%%%%%%%%%%%%%%%%%%%%%%%%%%
The initial optimal sensor/actuator placement is the assumed initial condition to begin the iterative sensor/actuator placement, $\mathcal{C}_0 = \mathcal{C}^\star$, and data-driven array modeling is implemented to converge on the final design in an iterative loop as shown in Algorithm~\ref{alg:dmd_placement}:  
\begin{algorithm}
	\caption{Iterative DMD-Based Sensor/Actuator Placement Optimization}
	\label{alg:dmd_placement}
	\begin{algorithmic}[1]%[h!]
		\REQUIRE Time-series data $\mathbf{X} \in \mathbb{R}^{n \times m}$, initial sensor location $\mathcal{C}_0$, convergence tolerance $\epsilon$
		\ENSURE Optimal sensor locations $\mathcal{C}^\star$ by minimizing cost $J(\mathcal{C}^\star)$
		\STATE Determine unloaded array dynamics 
		\STATE Construct Hankel matrix $\mathbf{H}^{(0)}$ from unloaded array data
		\STATE Perform DMD on $\mathbf{H}^{(0)}$ to obtain modes $\boldsymbol{\Phi}^{(k)}$ and eigenvalues $\boldsymbol{\Lambda}^{(k)}$
		\STATE Evaluate Maghami-Joshi cost function for optimal placement: $J(\mathcal{C}) = \sum_{i=1}^{n_r} 1 / \sigma_i$
		\STATE Set iteration counter $k = 1$
		\REPEAT
		\STATE Model array with current placement loads 
		\STATE Construct Hankel matrix $\mathbf{H}^{(k)}$ from loaded array data $p^{(k)}\in \mathcal{C}$
		\STATE Perform DMD on $\mathbf{H}^{(k)}$ to obtain modes $\boldsymbol{\Phi}^{(k)}$ and eigenvalues $\boldsymbol{\Lambda}^{(k)}$
		\STATE Evaluate Maghami-Joshi cost function: $J(\mathcal{C})$
		\STATE Optimal sensor/actuator locations: $\mathbf{x}$
		\STATE Update system dynamics with new mass distribution
		\STATE $k = k + 1$
		\UNTIL $\mathcal{C}^{(k)} = \mathcal{C}^{(k-1)}$
		\RETURN $\mathcal{C}^* = \mathcal{C}^{(k)}$
	\end{algorithmic}
\end{algorithm}
Here, the array is theoretically loaded by the mass of the sensor/actuators. The ANC method in Eq.~\eqref{eq:ANC} yields adjusted mode shapes and associated frequencies, and the sensor/actuator optimal placement algorithm from \cite{Maghami1993} converges on a new placement. This placement reinitializes the loop by altering the mass properties, and the loop iterates in this fashion until sensor/actuator locations are converged upon. The objective function values from a final iteration of the loop are shown in Figure~\ref{fig:cand2}.
% \medskip
%%%%%%%%%%%%%%%%%%%
\section{Results}
%%%%%%%%%%%%%%%%%%%
The sensor/actuator placement optimization demonstrated in this manuscript maintains that the reduced-order models allow controllability and observability of key dynamic characteristics. The result of the analysis above is implemented in a simplistic Linear Quadratic Regulator (LQR) formulation to ensure the analysis prediction is aligned with the optimized system of interest, whose sensor/actuator placement and mode frequencies are shown in Figure~\ref{fig:ANC2}.

\begin{figure}[tbh!]
\centering
\includegraphics[scale=0.6]{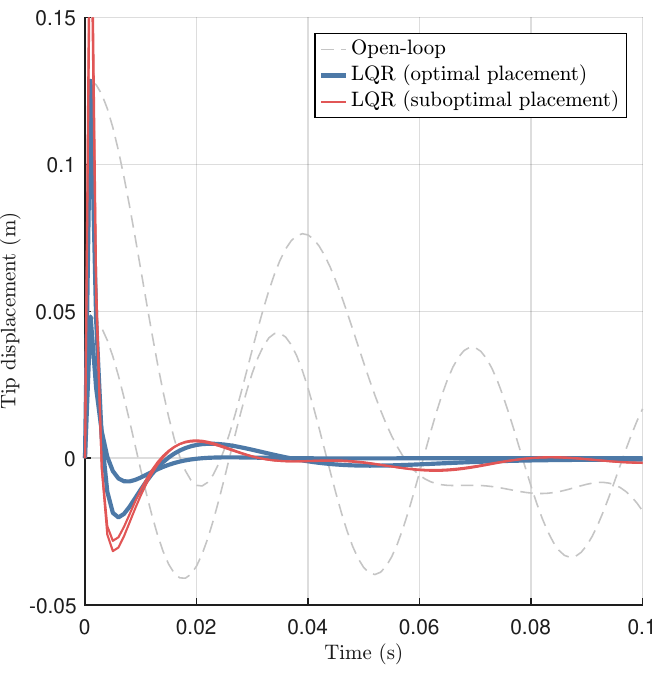}
\caption{\textbf{Open-loop, suboptimal closed-loop, and optimal closed-loop responses}}
\label{fig:LQR1}
% \end{figure}
%  \begin{figure}[tbh!]
\centering
\includegraphics[scale=0.6]{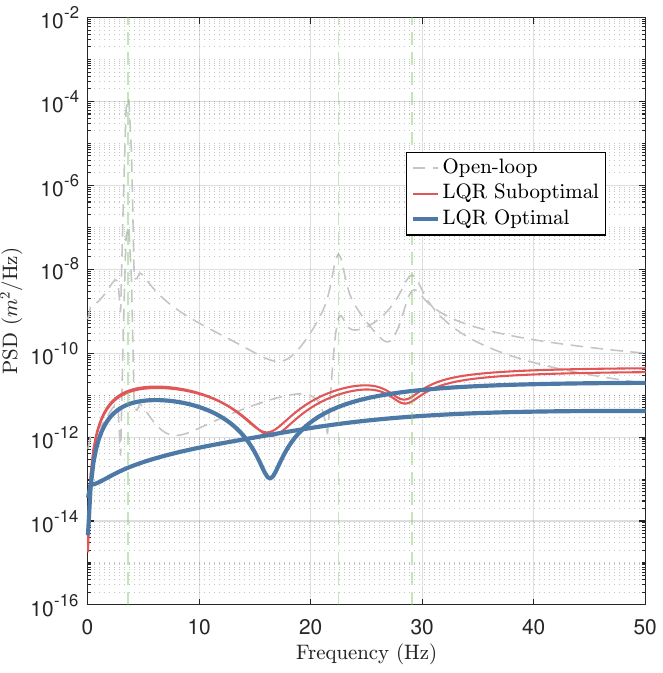}
\caption{\textbf{Open-loop, suboptimal closed-loop, and optimal closed-loop placement output PSD}}
\label{fig:PSD}
 \end{figure}

The LQR was constructed with the intent of suppressing the vibration modes intrinsic to the solar array. The vibration-suppression objective was first formulated in modal coordinates, allowing each flexible mode to be individually damped through feedback. Controller design was achieved through recasting the modal formulation into a Linear Time Invariant (LTI) system, such that modal dynamics were preserved and represented. In the LTI formulation, the control input matrix $B$ and the output matrix $C$ serve as direct surrogates for actuator and sensor placement. Since $B$ and $C$ were selected via Hankel-based optimization, the closed-loop system was expected to suppress vibrational modes with the minimum energy output. In contrast, when suboptimal placement is enforced, the control input or resulting feedback degrades vibration suppression. In Figure~\ref{fig:LQR1}, the time-domain simulation of a step response triggers vibration in the dominant modes, shown as a black dashed line. This is the open-loop response of the system. The observation and control of the system at the optimal node points is simulated to show that the response can be arrested, and the controlled response for both sensor/actuator pairs in the optimal configuration is shown in blue. As a point of comparison, the sensor/actuator locations initially converged upon before the modified dynamics had been taken into account; the suboptimal placements were assessed by the same LQR control scheme, and the time domain response of the pair of sensors/actuators is shown in red.
This time-domain reconstruction provides confidence that the optimally selected locations ensure observability and controllability of the modal system, beyond the intuition that the theoretical underpinnings offer. Furthermore, checking the frequency domain characteristics of the system allows another assurance that the open-loop behavior is nulled by the optimal sensor/actuator locations more efficiently than other locations. In the open loop Power Spectral Density plot (PSD), Figure~\ref{fig:PSD}, the system mode peaks are prominent due to the nature of an impulse response on lightly damped modes. In the closed-loop PSD, the controller should reduce the height of these peaks or eliminate them entirely, suggesting less energy in each mode. Overall, the lowest integrated PSD value (the least area under the PSD curve) suggests the closed-loop system that controls the dynamic response best. In Figure~\ref{fig:PSD}, it is clear by visual inspection that the PSD of the optimal closed-loop system has the lowest area under the curve. Quantitative results from these comparisons are summarized in Table \ref{tab:LQRresults}. The metrics include integrated PSD values, step-response performance (overshoot and settling time), and total control effort. The table values provide a quantitative performance comparison of the optimal system. The optimal placement yields the lowest vibration energy, fastest settling, and lowest control effort, which is what was set out to be proven.
\begin{table}[t]
	\renewcommand{\arraystretch}{1.5}
	\caption{Performance metrics comparing optimal, suboptimal, and open-loop sensor/actuator configuration}
	\centering
	\resizebox{\columnwidth}{!}{
	\begin{tabular}{lrrr}
		\hline
		\textbf{Metric} & \textbf{Channel 1} & \textbf{Channel 2} & \textbf{Aggregate} \\
		\hline
		\multicolumn{3}{l}{\textit{Variance from Spectrum (Integrated PSD)}}&  $(\Sigma)$\\
		Optimal   & $2.566\times10^{-8}$ & $3.744\times10^{-9}$ & $2.940\times10^{-8}$ \\
		Suboptimal& $7.189\times10^{-8}$ & $5.916\times10^{-8}$ & $1.310\times10^{-7}$ \\
		Open-loop &$ 2.219\times10^{-4}$ & $5.748\times10^{-6}$ & $2.276\times10^{-4}$ \\
		\hline
		\multicolumn{3}{l}{\textit{Overshoot (\%)}}& (average) \\
		Optimal   & 17.84 & 6.34 & 12.09 \\
		Suboptimal& 29.05 & 25.94 & 27.49 \\
		Open-loop & 56.39 & 27.29 & 41.84 \\
		\hline
		\multicolumn{3}{l}{\textit{Settling Time (s)}} & (max) \\
		Optimal   & 0.126 & 0.060 & 0.126 \\
		Suboptimal& 0.153 & 0.152 & 0.153 \\
		Open-loop & $>$10 & $>$10 & $>$10 \\
		\hline
		\multicolumn{3}{l}{\textit{Control Effort ($\int u^2 dt$)}}&  $(\Sigma)$ \\
		Optimal   &$5.447\times10^{-4}$ &$2.916\times10^{-4}$ &$8.363\times10^{-4}$ \\
		Suboptimal&$6.397\times10^{-4}$ &$6.703\times10^{-3}$ &$1.310\times10^{-3}$ \\
		\hline
	\end{tabular}}
	\label{tab:LQRresults}
\end{table}	
%%%%%%%%%%%%%%%%%%%%%
\section{Conclusion}
%%%%%%%%%%%%%%%%%%%%%
Throughout this paper, a data-driven approach has been systematically defined and verified that optimizes the placement of sensor/actuator pairs on a flexible structure. This approach was applied over a definable and characterizable system to ensure a comparison of the method and proof of its validity. First, DMD was shown to provide an efficient method for system reduction and dominant dynamics extraction, verified by a frequency domain analysis. Next, optimal locations of the sensor/actuators were derived from the data-driven Hankel matrix, for all combinations of sensor locations. The perturbation of system mass properties from sensor/actuator placement was integrated to ensure representative dynamics. This process was iterated to converge on optimal sensor placement driven by modified dynamics from those sensors. Finally, the sensor/actuator placement was implemented in an LQR optimal control scheme set to null vibration response, and results proved that the optimized locations required the least energy and showed the best time-domain performance. Overall, a method was proven that is scalable to highly complex systems that do not have analytical solutions from which to derive sensor/actuator placement methods. Future work will consider LQR controller performance as an inner optimization loop, thus establishing a joint design problem that treats sensing, actuation, and control as coupled decisions.

%\cite{BruntonKutz2019}

%%%%%%%%%%%%%%%%%%%%%%%%%%%%%%%%%%%%%%%%%%%%%%%%%%%%%%%%%%%%%%%%%%%%%%%%%%%%%%%%%%%%%%%%%%%%%%%%%%%%%%
\bibliographystyle{IEEEtran}
\bibliography{references}

\thebiography
%% This biostyle allows you to insert your photo size 1in X 1.25in
\begin{biographywithpic}{Matthew Hilsenrath}{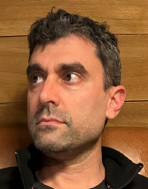}
is a Senior Staff Engineer with the Lockheed Martin Advanced Technology Center. He has made contributions in National Security Space, Strategic Defense, and Human Spaceflight over his career. He has a bachelor's degree in Aerospace Engineering from the University of Florida, a master's degree in Space Systems Engineering from the Johns Hopkins University, and is a PhD candidate in Systems Engineering with a focus in combined physical and control system design (control co-design) at Colorado State University. His interests include data-driven control design, non-deterministic error analysis, and digital signal processing.
\end{biographywithpic} 

\begin{biographywithpic}{Daniel R. Herber}{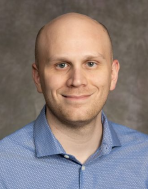}
is an Associate Professor in the Department of Systems Engineering at Colorado State University in Fort Collins, CO, USA. His research interests and projects have been in design optimization, model-based systems engineering, system architecture, digital engineering, dynamics and control, and combined physical and control system design (control co-design), frequently collaborating with academia, industry, and government laboratories. His work has involved several application domains, including energy, aerospace, defense, and software systems. He teaches courses in model-based systems engineering, system architecture, controls, and optimization. He is a member of INCOSE, ASME, and AIAA.
\end{biographywithpic}

% \clearpage

% \printinunitsof{in}

% textwidth = \prntlen{\textwidth}

% columnwidth = \prntlen{\columnwidth}

\end{document}